# Optimal Design of Layered Periodic Composites for Mitigation of Impact-Induced Elastic Waves

H. Sadeghi[1], S. Nemat-Nasser

Center of Excellence for Advanced Materials, Department of Mechanical and Aerospace Engineering,

University of California San Diego, CA, USA 92093-0416

**ABSTRACT**

A systematic method for optimal design of layered periodic composites for mitigation of impact-induced shock waves is presented. Frequency spectrum of a pulse with a sharp rise-time is analyzed and the frequency range that carries most of the pulse energy is identified. A genetic algorithm is used to maximize the stop bands of a layered periodic composite over the target frequency range. Due to reflection of the pulse over the stop bands, the maximum stress and the energy of transmitted pulse become minimal. To verify the theoretical calculation a sample is fabricated and Hopkinson bar experiments are performed. It is observed that only 9.7% of energy of the incident pulse gets transmitted through the sample. In addition, the wave speed in the composite is measured to be 45.4% less than the wave speed in its constituent material with the lowest wave speed.

## 1 INTRODUCTION

In recent years, periodic composites (PCs) have attained considerable attention because of their interesting elastodynamic behavior [**1**,**2**,**3**,**4**]. These composites display stop bands where an incident elastic wave gets effectively reflected. This feature can be used to design acoustic filters, noise insulators, and vibrationless environment to house sensitive instruments. Different methods have been used to design PCs for a desired elastodynamic response [**5**,**6**,**7**]. Hussein et al [**5**] used a multiobjective genetic algorithm for design of layered PCs for an optimal frequency band structure. They illustrated examples for optimizing the performance of layered PCs for acoustic filtering. Meng et al [**6**] performed optimization to enhance the underwater sound absorption of an acoustic metamaterial slab. They illustrated the feasibility of combining several layers of different resonant frequencies to achieve a broadband underwater sound absorber. Wang et al [**7**] presented a method for design of layered elastic metamaterials for maximizing frequency range where a PC exhibits negative effective elastodynamic properties. They used genetic algorithm for optimization and showed that their approach provides satisfactory results.

---

[1] hsadeghi@ucsd.edu

There have been many efforts to understand the behavior of layered materials under high strain rate loading. Zhuang et al [8] investigated finite amplitude shock propagation in layered periodic composites experimentally. They observed that due to scattering effects, the shock speed in the composite is lower than the shock speed in its constituent materials. Chen and Chandra [9] considered the effect of heterogeneity in layered composites under impact loading. They studied the effect of impedance mismatch, thickness ratio, and interface density on the response of layered composites. Luo et al. [10] studied stress wave propagation in two and three layers structures under impulsive loading. They showed that when an incident pulse passes through a layered structure, a reduced stress amplitude and elongated pulse duration can be obtained with proper selection of materials and layers dimensions. Schimizze et al [11] studied blast induced shock wave mitigation in sandwich structures. They observed that the density and acoustic impedance mismatch are of primary importance in shock wave mitigation in this kind of structures. Due to promising behavior of layered structures under high strain rate loading, it is necessary to develop design methods for their optimal performance.

In this paper, a systematic method for optimal design of layered PCs for mitigation of impact-induced elastic waves is presented. Frequency spectrum of a pulse with a sharp rise time is studied and the frequency range that contains most of the pulse energy is found. A genetic algorithm is used to design the topology of a layered PC to maximize the stop bands over the target frequency range. A constraint is introduced into the optimization problem to limit the total unit cell size of the PC. Transfer matrix calculations and Hopkinson bar experiments are performed to study the behavior of the composite subjected to the pulse.

## 2 WAVE PROPAGATION IN LAYERED PERIODIC COMPOSITES

### 2.1 Band structure calculation

Consider a unit cell of an infinite layered PC consisting of $N$ different homogeneous layers, as shown in Figure 1. In this Figure, $E^{(j)}, \rho^{(j)}$, and $d^{(j)}$ are the elastic modulus, density, and thickness of the $j$-th layer, respectively. The displacement, $u$, and stress, $\sigma$, at the left boundary of the first layer in the unit cell, $x^{1L}$, can be related to those at the right boundary of the $N$-th layer, $x^{NR}$, by

$$\begin{bmatrix} u(x^{NR}) \\ \sigma(x^{NR}) \end{bmatrix} = \mathbf{T} \begin{bmatrix} u(x^{1L}) \\ \sigma(x^{1L}) \end{bmatrix} \quad (1)$$

where $\mathbf{T} = \mathbf{T}_N \mathbf{T}_{N-1} \dots \mathbf{T}_1$ is the transfer matrix of the unit cell and $\mathbf{T}_j$ is the transfer matrix of the $j$-th layer given as

$$\mathbf{T}_j = \begin{bmatrix} \cos(k^{(j)} d^{(j)}) & \sin(k^{(j)} d^{(j)})/Z^{(j)} \\ -Z^{(j)} \sin(k^{(j)} d^{(j)}) & \cos(k^{(j)} d^{(j)}) \end{bmatrix}. \quad (2)$$

where $Z^{(j)} = \rho^{(j)} c^{(j)2} k^{(j)}$, $\omega$ is the angular frequency, $k^{(j)}$ is the wave number in the $j$-th layer, and $c^{(j)}$ is the wave velocity in the $j$-th layer. For Bloch type waves stress and displacement at the boundaries of the unit cell are related by

$$\begin{bmatrix} u(x^{NR}) \\ \sigma(x^{NR}) \end{bmatrix} = e^{iQ} \begin{bmatrix} u(x^{1L}) \\ \sigma(x^{1L}) \end{bmatrix} \tag{3}$$

where $Q = kd$ is the normalized Bloch wave number. Combining eqs. (1) and (3) one can find the following eigenvalue problem

$$\mathbf{T}\mathbf{y}(x^{1L}) = e^{iQ}\mathbf{y}(x^{1L}) \tag{4}$$

which may be solved for either $\omega$ or $Q$ to find the band structure of the composite.

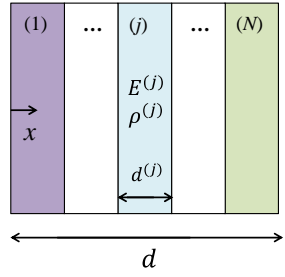

Figure 1. Unit cell of a layered periodic composite.

## 2.2 *Transmission and reflection calculation*

Consider a layered composite made of $m$ homogenous layers sandwiched between two homogenous semi-infinite bars as shown in Figure 2. Assume the amplitude of the incident wave in the incident bar is $A_+^{(0)}$, the amplitude of reflected wave in the incident bar is $A_-^{(0)}$, and the amplitude of wave transmitted to the transmission bar is $A_+^{(m+1)}$. Applying the continuity of displacement and stress at the interfaces of the composite with the incident and transmission bars, the reflection and transmission coefficients, $R$ and $T$, can be found as

$$R = \frac{A_-^{(0)}}{A_+^{(0)}} = -\frac{\mathbf{K}_{21}}{\mathbf{K}_{22}}, \quad T = \frac{A_+^{(m+1)}}{A_+^{(0)}} = \mathbf{K}_{11} - \frac{\mathbf{K}_{12}\mathbf{K}_{21}}{\mathbf{K}_{22}} \tag{5}$$

where $\mathbf{K} = \mathbf{L}_{m+1}^{-1}\mathbf{B}_{m+1}^{-1}\mathbf{T}'\mathbf{B}_0$, $\mathbf{T}' = \mathbf{T}_m \mathbf{T}_{m-1} \ldots \mathbf{T}_1$ is the transfer matrix of the composite, and $\mathbf{B}_j$ and $\mathbf{L}_j$ are given by

$$\mathbf{B}_j = \begin{bmatrix} 1 & 1 \\ iZ^{(j)} & -iZ^{(j)} \end{bmatrix}, \quad \mathbf{L}_j = \begin{bmatrix} \exp(ik^{(j)}x^{jL}) & 0 \\ 0 & \exp(-ik^{(j)}x^{jL}) \end{bmatrix} \tag{6}$$

The reflected energy, $Re$, and the transmitted energy, $Tr$, can therefore be calculated as

$$Re = RR^*, \quad Tr = TT^* \tag{7}$$

where $(.)^*$ denotes the complex conjugate.

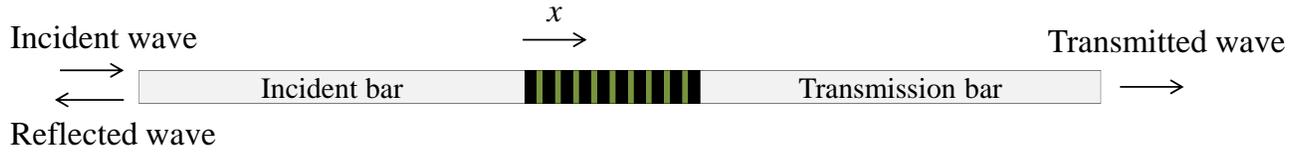

Figure 2. Finite slab of a layered PC sandwiched between two homogenous semi-infinite bars.

## 3  OPTIMIZATION

Here the goal is to find the topology of a layered PC to maximize the stop bands over a target frequency range, $f_1'$ to $f_2'$. The stop band ratio ($SR$) is defined as the ratio of sum of the stop bands frequencies (within the target frequency range) divided by the total frequency range

$$SR = \frac{\sum(f_2^{(i)} - f_1^{(i)})}{f_2' - f_1'} \qquad (8)$$

where $f_1^{(i)}$ and $f_2^{(i)}$ are the frequencies where the $i$-th stop band begins and ends, respectively. The material constituents of each phase are chosen based on the application and the thickness of each layer, $d^{(j)}$ ($j = 1, .., N$), are the design parameters. The design space for each design parameter is specified considering the frequency of interest, manufacturability, and other design limitations. Here, a constraint is introduced into the optimization problem to limit the total unit cell size of the composite to be equal to a prescribed value, $D$. The objective function is defined as the inverse of stop band ratio and the optimization problem can be stated as

$$\text{Minimize } \frac{f_2' - f_1'}{\sum(f_2^{(i)} - f_1^{(i)})}$$
$$\text{Subject to } \sum_{j=1}^{N} d^{(j)} = D \qquad (9)$$

This constrained optimization problem can be reduced to the following unconstrained one using quadratic penalty method [**12**]

$$\text{Minimize } F = \frac{f_2' - f_1'}{\sum(f_2^{(i)} - f_1^{(i)})} + \lambda \left(\sum_{j=1}^{N} d^{(j)} - D\right)^2 \qquad (10)$$

where $F$ is the new objective function and $\lambda$ is the penalty coefficient. Any global optimization method can be used to find the optimal solution. Here we adopt genetic algorithm (GA) [**13**] which is a global optimization method and search technique based on the principles of genetics and natural selection and has been used successfully in several engineering problems.

## 4  HOPKINSON BAR SETUP

Figure 3 shows a schematic representation of the Hopkinson bar setup used in this study. The striker bar hits the end A of the incident bar at a given velocity which produces a compressive pulse traveling along the bar. The sample is sandwiched between end B of the incident bar and end C of the transmission bar. When the pulse reaches the sample, part of it gets transmitted to the transmission bar and part of it gets reflected back into the incident bar. Strain gauges S1 and S2 measure the strain, $\varepsilon(t)$, in the middle of the incident and transmission bars as a function of time. Incident, transmission, and striker bars are made of steel with diameter of 0.5 in. The length of both incident and transmission bars is 4 ft, and the length of the striker bar is 4 in. The particle velocity, $V(t)$, and axial stress, $\sigma(t)$, in the bars can be calculated as [14]

$$V(t) = c_0 \varepsilon(t), \qquad \sigma(t) = \rho c_0 V(t) \tag{11}$$

where $c_0$ and $\rho$ are the wave speed and the density in the bars, respectively. Also, the total energy that pulse carries can be calculated as

$$E = e_0 \int_0^t V(\tau)^2 d\tau \tag{12}$$

where $e_0$ is a constant coefficient.

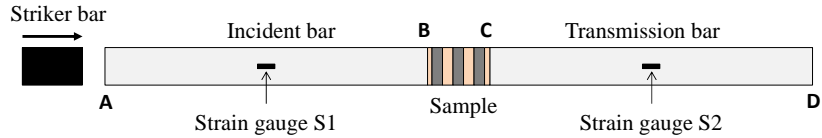

Figure 3. Schematic of the Hopkinson bar setup.

## 5  OPTIMAL DESIGN AND RESULTS

The incident pulse is measured in the Hopkinson bar by strain gauge S1 and its fast Fourier transform (FFT) is calculated. It is observed that a significant portion of the pulse energy is carried by the components with frequencies over 0 to 100 kHz. Therefore, the target frequency range is chosen to be 0 to 100 kHz to maximize the stop bands. The design space is $0.1\ cm < d^{(j)} < 1\ cm$ for thickness of each layer and the total unit cell size is $D = 2\ cm$. Here 2-phase composites with periodic layers of polycarbonate and steel are considered and the number of layers in the unit cell, $N$, can change (see Figure 4). Without loss of generality, it is assumed that the first layer in the unit cell is made of polycarbonate. The material properties of each phase are given by $c_{st} = 5064\ m/s$, $\rho_{st} = 7810\ kg/m^3$, $c_{pc} = 2236\ m/s$, and $\rho_{pc} = 1193\ kg/m^3$, for the wave speed and density of steel and polycarbonate, respectively. Table 1 shows the optimal design and stop band ratio for different values of $N$. It can be seen that $N = 3$ gives the largest stop band ratio and it is chosen as the optimal design. It is understood that in this case increasing the number of layers does not necessarily produce wider stop bands. Figure 5(a) shows the corresponding band structure for the optimal design. It can be seen that there is a wide stop band from 28 to 104 kHz. Figure 5(b) shows the transmission and reflection spectra of 5 unit cells of the composite sandwiched between two semi-infinite steel bars. It can be seen that components of the pulse with frequency content over the stop bands are completely reflected. Also, a significant portion of the waves with frequencies over the pass bands get reflected as well.

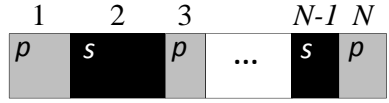

Figure 4. Unit cell of a 2-phase layered PC (*p*: polycarbonate; and *s*: steel).

Table 1. Optimal design for polycarbonate/steel layered PC

| N | $d^{(j)}$ (for $j = 1..N$) (mm) | SR |
|---|---|---|
| 2 | 14.6, 5.4 | 59.5 |
| 3 | 4.0, 9.6, 6.4 | 73.2 |
| 4 | 1.1, 3.9, 8.7, 6.3 | 61.5 |
| 5 | 2.2, 7.3, 4.2, 1.1, 5.2 | 62.1 |

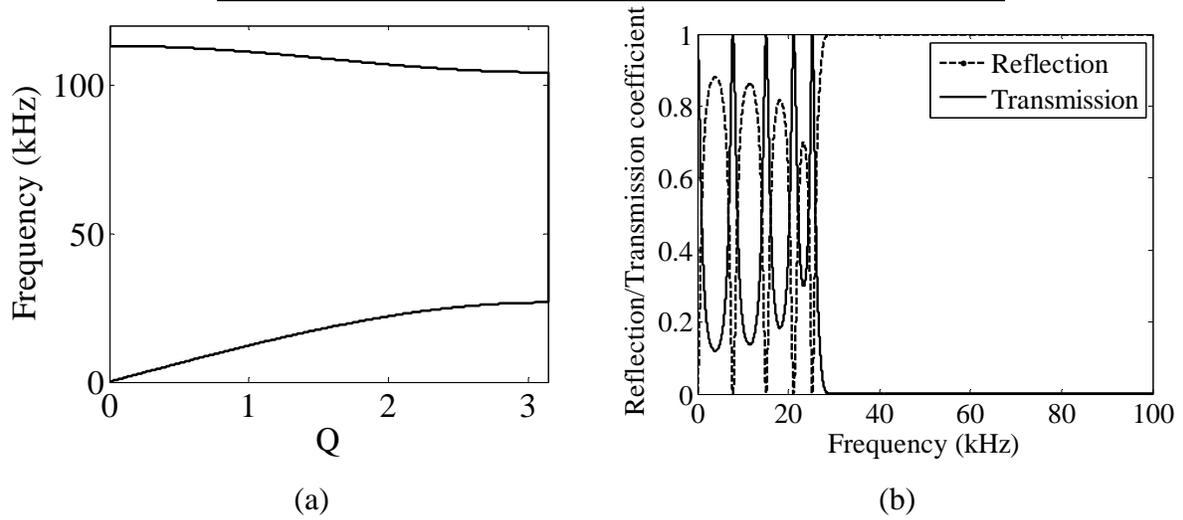

Figure 5. (a) Band structure for the optimal design ($N = 3$); and (b) reflection and transmission spectra of the sample sandwiched between two semi-infinite steel bars

Figure 6(a) shows the transmitted pulse measured experimentally together with the incident pulse as a function of time (pulses are shifted in time for better comparison). It can be seen that the rise time of transmitted pulse is $\tau_{tr}^* = 62.5~\mu s$ which is 4.3 times larger than the incident pulse rise time, $\tau_{in}^* = 14.5~\mu s$. Also, the maximum stress of the transmitted pulse is $\sigma_{tr}^* = 37.1~MPa$ which is 3.4 times less than that of the incident pulse, $\sigma_{in}^* = 125.2~MPa$. Figure 6(b) shows FFT of the incident and transmitted pulses. It can be seen that components of the pulse with frequencies above 28 kHz, which are within the stop bands of the composite, are not transmitted. Increase in the rise time and decrease in the maximum stress of the transmitted pulse compared to those of the incident pulse are due to: (i) significant portion of the incident pulse being within the stop bands and getting reflected, and (ii) multiple reflections happening at the interfaces of the composite for the parts of the pulse within the pass bands. Furthermore, the wave speed in the sample is measured to be 1538 $m/s$ which is 45.4% less than the wave speed in polycarbonate. Also, the energy of the incident and the transmitted pulses are calculated and it is observed that only 9.7% of the incident pulse energy gets transmitted.

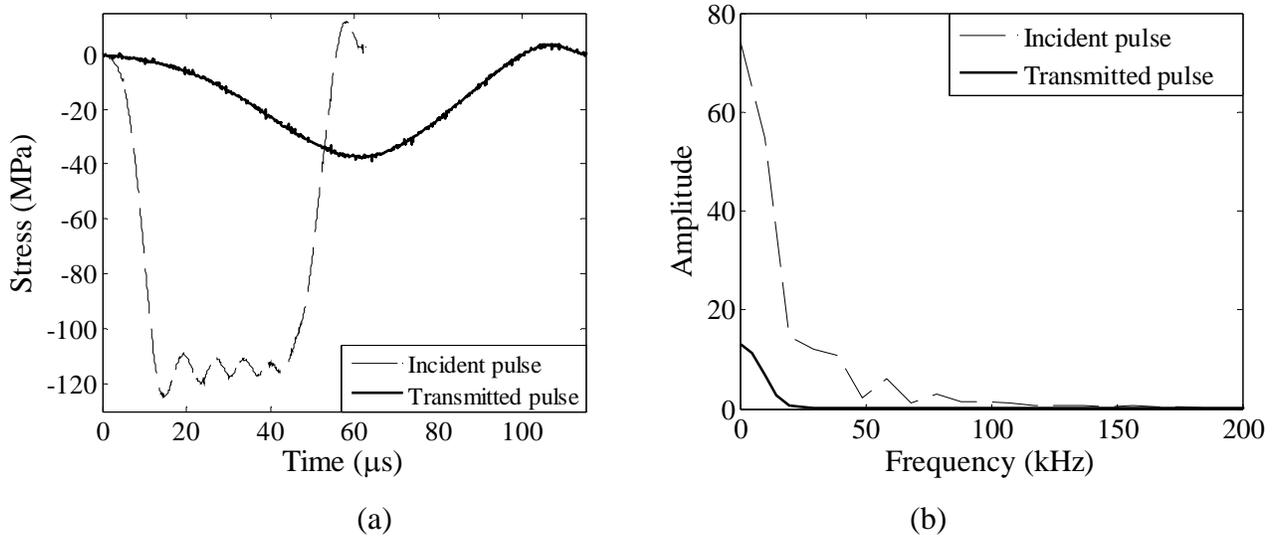

Figure 6. (a) Experimentally measured incident and transmitted pulses; and (b) FFT of the incident and transmitted pulses.

## 6  CONCLUSION

A systematic method for optimal design of layered periodic composites for mitigation of impact-induced shock waves is presented. Frequency spectrum of a pulse with a sharp rise time is studied and the frequency range that contains most of the pulse energy. Genetic algorithm is used to design a layered periodic composite to maximize the stop bands over the target frequency range. Components of the pulse with frequency content within the stop bands of the composite get effectively reflected, which results in the maximum stress and the energy of transmitted pulse become minimal. Hopkinson bar experiments are performed to study the behavior of the sample subjected to the pulse. Experimental results show that only 9.7% of energy of the incident pulse gets transmitted through the sample.

## 7  ACKNOWLEDGEMENT

This research has been conducted at the Center of Excellence for Advanced Materials (CEAM) at the University of California, San Diego, under DARPA Grant RDECOM W91CRB-10-1-0006 to the University of California, San Diego.

## 8  REFERENCES

[1] Alireza V. Amirkhizi and Sia Nemat-Nasser, "Experimental verification of stree-wave bands and negative phase velocity in layered media," *Theoretical Applied Mechanics*, vol. 38, no. 4, pp. 299-319, 2011.

[2] S. Nemat-Nasser, J.R. Willis, A. Srivastava, and A.V. Amirkhizi, "Homogenization of periodic elastic composites and locally resonant sonic materials," *Physical Review B*, vol. 83, no. 10, p. 104103, 2011.

[3] Sia Nemat-Nasser, Hossein Sadeghi, Alireza V Amirkhizi, and Ankit Srivastava, "Phononic Layered Composites for Stress-wave Attenuation," *Submitted to Mechanics Research Communications*.

[4] S Nemat-Nasser, FCL Fu, and S Minagawa, "Harmonic waves in one-, two-and three-dimensional composites: bounds for eigenfrequencies," *International Journal of Solids and Structures*, vol. 11, no. 5, pp. 617-642, 1975.

[5] Mahmoud I Hussein, Karim Hamza, Gregory M Hulbert, Richard A Scott, and Kazuhiro Saitou, "Multiobjective evolutionary optimization of periodic layered materials for desired wave dispersion characteristics," *Structural and Multidisciplinary Optimization*, vol. 31, no. 1, pp. 60-75, 2006.

[6] Hao Meng, Jihong Wen, Honggang Zhao, and Xisen Wen, "Optimization of locally resonant acoustic metamaterials on underwater sound absorption characteristics," *Journal of Sound and Vibration*, vol. 331, no. 20, pp. 4406-4416, 2012.

[7] Wei Wang and Tianzhi Yang, "Multi-Objective Optimization of Layered Elastic Metamaterials With Multiphase Microstructures," *Journal of Vibration and Acoustics*, vol. 135, no. 4, p. 041010, 2013.

[8] Shiming Zhuang, Guruswami Ravichandran, and Dennis E Grady, "An experimental investigation of shock wave propagation in periodically layered composites," *Journal of the Mechanics and Physics of Solids*, vol. 51, no. 2, pp. 245-265, 2003.

[9] X Chen and Namas Chandra, "The effect of heterogeneity on plane wave propagation through layered composites," *Composites science and technology*, vol. 64, no. 10, pp. 1477-1493, 2004.

[10] Xiaobo Luo, Amjad J Aref, and Gary F Dargush, "Analysis and optimal design of layered structures subjected to impulsive loading," *Computers & Structures*, vol. 87, no. 9, pp. 543-551, 2009.

[11] Benjamin Schimizze, Steven F Son, Rahul Goel, Andrew P Vechart, and Laurence Young, "An experimental and numerical study of blast induced shock wave mitigation in sandwich structures," *Applied Acoustics*, vol. 74, no. 1, pp. 1-9, 2013.

[12] David G Luenberger, *Introduction to linear and nonlinear programming*.: Addison-Wesley Reading, MA, 1973, vol. 28.

[13] Randy L Haupt and Sue Ellen Haupt, *Practical genetic algorithms*.: John Wiley & Sons, 2004.

[14] Sia Nemat-Nasser, Jon B Isaacs, and John E Starrett, "Hopkinson techniques for dynamic recovery experiments," *Proceedings of the Royal Society of London. Series A: Mathematical and Physical Sciences*, vol. 435, no. 1894, pp. 371-391, 1991.